\documentclass[aps,prd, reprint, superscriptaddress]{revtex4-1}
\usepackage{natbib}
\usepackage[pdftex]{graphicx}
\usepackage{color}
\usepackage{amsmath}
\usepackage{rotating}
\usepackage{wasysym}
\usepackage{cancel}

\def\unitn{Dipartimento di Fisica, Universit\`a di Trento and INFN,
Gruppo Collegato di Trento, 38123 Povo, Trento, Italy}

\def\esac{SRE-OD ESAC, European Space Agency, Camino bajo del Castillo s/n,
Urbanizaci\'on Villafranca del Castillo, Villanueva de la Can\~ada, 28692 Madrid, Spain}

\def\aei{Albert-Einstein-Institut, Max-Planck-Institut f\"ur Gravitationsphysik und Universit\"at Hannover,
Callinstrasse 38, 30167 Hannover, Germany}

\def\apc{APC, Universit\'e Paris Diderot, CNRS/IN2P3, CEA/Ifru, Observatoire de Paris, Sorbonne Paris Cit\'e, 10 Rue A.\,Domon et L.\,Duquet, 75205 Paris Cedex 13, France}

\def\ieec{Institut de Ci\`encies de l'Espai, (CSIC-IEEC), Facultat de Ci\`encies,
Campus UAB, Torre C-5, 08193 Bellaterra, Spain}

\def\ethz{ETH Z\"urich, Institut f\"ur Geophysik, Sonneggstrasse 5, 8092 Z\"urich}

\def\estec{European Space Technology Centre, European Space Agency, Keplerlaan 1, 2200 AG Noordwijk, The Netherlands}

\def\goddard{Gravitational Astrophysics Lab, NASA Goddard Space Flight Center}

\def\icl{High Energy Physics Group, Imperial College London. Blackett Laboratory, Prince Consort Road, London, SW7 2AZ}

\begin{document}

\title{Optimal design of calibration signals in space borne gravitational wave detectors}

\author{Miquel~Nofrarias}
\email{nofrarias@ice.cat}
\affiliation{\ieec}

\author{Nikolaos~Karnesis}\affiliation{\ieec}
\author{Ferran~Gibert}\affiliation{\ieec}
\author{Michele~Armano}\affiliation{\esac}
\author{Heather~Audley}\affiliation{\aei}
\author{Karsten~Danzmann}\affiliation{\aei}
\author{Ingo~Diepholz}\affiliation{\aei}
\author{Rita~Dolesi}\affiliation{\unitn}
\author{Luigi~Ferraioli}\affiliation{\ethz}
\author{Valerio~Ferroni}\affiliation{\unitn}
\author{Martin~Hewitson}\affiliation{\aei}
\author{Mauro~Hueller}\affiliation{\unitn}
\author{Henri~Inchauspe}\affiliation{\apc}
\author{Oliver~Jennrich}\affiliation{\estec}
\author{Natalia~Korsakova}\affiliation{\aei}
\author{Paul~W.~McNamara}\affiliation{\estec}
\author{Eric~Plagnol}\affiliation{\apc}
\author{James~I.~Thorpe}\affiliation{\goddard}
\author{Daniele~Vetrugno}\affiliation{\unitn}
\author{Stefano~Vitale}\affiliation{\unitn}
\author{Peter~Wass}\affiliation{\icl}
\author{William~J.~Weber}\affiliation{\unitn}

% Date
\date{\today}

\begin{abstract}
Future space borne gravitational wave detectors will require 
a precise definition of calibration signals to ensure the 
achievement of their design sensitivity. 
The careful design of  the test signals plays a key role in the correct 
understanding and characterisation of these instruments. In that sense, methods achieving 
optimal experiment designs must be considered as complementary to the parameter estimation methods being used to determine the parameters describing the system. 
The relevance of experiment design is particularly significant for the 
LISA Pathfinder mission, which will spend most of its operation time performing experiments 
to characterise key technologies for future space borne gravitational wave observatories. 
Here we propose a framework to derive the optimal signals ---in terms of 
minimum parameter uncertainty--- to be injected to these instruments during its calibration 
phase. We compare our results with an alternative numerical algorithm which achieves an optimal 
input signal by iteratively improving an initial guess. We show agreement of both 
approaches when applied to the LISA Pathfinder case.
\end{abstract}

%\keywords{}

\maketitle
%\tableofcontents

\section{Introduction}
\label{sec:intro}

LISA Pathfinder~\cite{Antonucci11_labExp} is an ESA mission with NASA contributions designed to test key technologies for the detection of gravitational waves in space, like the proposed eLISA~\cite{eLISA}. The main scientific goal for the mission is 
expressed in terms of a differential acceleration noise between two test masses in nominally geodesic 
motion down to a level of $\rm S_{\Delta a} = 3 \times 10^{-14}\,m/s^2/\sqrt{Hz}$ at 3\,mHz. 
The relevance of this requirement is not only its demand in terms of  noise reduction but also the 
very low frequency measuring band, which introduces technological difficulties that can not be addressed
by ground based gravitational wave detectors due to the so called seismic wall~\cite{Abadie10}.

The LISA Pathfinder mission is currently planned to have a six month operations period at the 
Lagrange point L1 that will be split between the two experiments on-board: the European 
LISA Technology Package (LTP) and the American Disturbance Reduction System (DRS).
This leads to a very short operation period of roughly three months for the complete characterisation
and achievement of the scientific goal for the LTP. 

It is worth noticing that, after the demonstration of the technology readiness, a second ---yet not less relevant-- objective of the mission is a detailed characterisation of the noise contributions to the main scientific measurement. An extensive list of experiments has been put forward by the scientific team including 
experiments to characterise the optical metrology~\cite{Audley11}, the inertial sensor instrument~\cite{Dolesi03}, the effects of the thermal~\cite{Canizares09} and magnetic~\cite{Diaz-Aguilo12} environment, and pure free-fall experiments that aim to measure acceleration noise in an configuration that is even more representative of eLISA~\cite{Grynagier09}. All these runs need to be executed
via tele-commands using a daily 8\,hours communication window with the satellite. Internal constraints in 
pre-processing and validation of tele-commands will add a latency from 2 to 3 days between the definition of a tele-command sequence and its execution on the spacecraft.

The planning of experiments represents therefore a crucial part of the mission and 
needs to be optimised 
accordingly to make sure that the information obtained from each experiment is maximised. 
As part of this effort a MATLAB toolbox has been developed with the specific aim to deal with 
the LTP data during flight operations~\cite{Hewitson09}. Among the different methods and capabilities of this tool, much attention has been paid to the improvement of the methods to obtain precise parameters 
from the experiments~\cite{Nofrarias10,Ferraioli11,Congedo12, Karnesis14, Vitale14}.
These have been tested with simulated data, taking into account the expected noise 
performance of the Pathfinder mission, in a series of mock data challenges with data generating using the analysis software's built-in modeling and simulation tools. Agreement
between methods was also checked with data generated from an independent spacecraft simulator developed by the prime industrial contractor, as was the case in the LISA Pathfinder operational exercises~\cite{Nofrarias12}.

These analyses focused on the parameter estimation strategy and the achievement 
of an optimal precision in the parameters obtained, following the heritage of previous simulated 
data exercises, like for instance the LISA Mock Data Challenge~\cite{Babak10} that focused on the problem of astrophysical parameter estimation from LISA data. Unlike the problem of astrophysical data analysis, in the LISA Pathfinder case, the measured signal is the response of the LISA Pathfinder system to some injected input signal that was specified by the telecommand file. In other words, there exists the opportunity in LISA Pathfinder to design the injected signals so that the measurement of the system parameters is optimised. The operators of ground-based gravitational-wave detectors have a similar opportunity to design signals when characterising the response of their instruments to various noise sources but, given their easy access to their instruments, not as much emphasis is placed on optimising signal injections. 
Instead, for a space borne gravitational wave observatory, such optimal experiment designs might prove very important for maximizing science return for a given mission duration.
LISA Pathfinder thus represents a scenario where careful signal design would produce the most benefit. In the following we propose a general framework which allows the optimisation of the input signals applied to a given system. 

Optimal experiment design~\cite{Fedorov, Goodwin, Pukelsheim} is a long-standing area of research. In general terms, the main objective is to adjust the experiment in such a way that the maximal information is obtained from the data. This general purpose has of course applicability in a wide variety of areas spanning the study of physical, biological or engineering systems. The reader is referred to reviews covering this extensive field of research for more insight~\cite{Mehra74, Walter90, Chaloner95, Gevers05, Hjalmarsson05}.  
In most cases, experiment design is described as an optimisation problem 
for a given figure of merit, which typically relates to a scalar of the Fisher information matrix.
Although the description used here applies to a general case, in the current work we will be mostly interested in the application to the estimation of the main parameters governing the combined dynamics of the test mass and the spacecraft in LISA Pathfinder.
Hardware on-board the satellite imposes us a further limitation which is only
to consider sinusoidal signals as input signals.

This work is organised as follows. In section~\ref{sec.definition} we introduce the problem of 
experiment design and the notation used in this work. Section~\ref{sec.design} describes
a numerical algorithm to optimise the signal to be injected given a model, and its application 
to a simple case. In section~\ref{sec.LPFmodel} we introduce the LISA Pathfinder model 
used for our analysis in section~\ref{sec.LPFOptSig} and finally present our conclusions.

\section{Fisher matrix analysis \label{sec.definition}}

\subsection{Definitions and notation}

In the following we will describe a given system as
\begin{equation}
\vec{o}(\omega) = {\bf H}(\omega; \Theta)\,\vec{s}(\omega) + \vec{n}(\omega)
\label{eq.system}
\end{equation}
where $\vec{o}$ is a vector with the measurements being considered, $\vec{s}$ is a vector with injection signals that can be applied to test the system and ${\bf H}(\omega; \Theta)$ is the matrix whose components, $H_{\rm ij}(\omega; \Theta)$, 
contain the transfer function describing the dynamics of the system in the frequency domain with a 
dependence on a set of parameters $\Theta = \lbrace \theta_1, \cdots, \theta_N\rbrace$. 
$\vec{n}(\omega)$ describes the noise contribution of our instrument.
% Analogously, ${\bf G}(z; \Theta)$ contains the parameter dependence and dynamics of the noise contributions.

The likelihood function is the probability to observe a measurement for a given set of parameters describing that system. 
Assuming that the data is Gaussian distributed, the likelihood for our system will be
\begin{equation}
p(\vec{o}\,| \Theta)  = \newline
 [2\pi {\bf \Sigma}]^{-1/2} \exp[ -\frac{1}{2} (\vec{o} - {\bf H}(\Theta)\cdot \vec{s}){\bf \Sigma^{-1} }
(\vec{o} - {\bf H}(\Theta)\cdot \vec{s})]   
\end{equation}

where $\bf \Sigma$ is the noise covariance matrix. Experiment design is based on the analysis of the Fisher matrix, whose elements are defined as
\begin{equation}
F_{ij} = \left. \left< \left(\frac{\partial \log(p(\vec{o}\,|\Theta))}{\partial \Theta_i} \right)^T  
\left(\frac{\partial \log(p(\vec{o}\,| \theta))}{\partial \theta_j}\right) \right> \right|_{\theta_0}
\label{eq.Fisher}
\end{equation}
which can be used to set limit for expected covariance matrix of the parameters, know as the Cr\'amer-Rao bound~\cite{Vallisneri08}
\begin{equation}
 {\rm cov}[\theta_i,\theta_j] \ge {\bf F^{-1}}
\end{equation}

%\subsection{Eigendecomposition of the Fisher matrix}
The decomposition of the Fisher matrix into eigenvalues and eigenvectors will prove to be very 
useful in the following sections. Given a N$\times$N Fisher matrix {\bf F}, defined by a set of N parameters,
the eigenvectors $\vec{u}$ and eigenvalues, $\lambda$, always fulfil 
\begin{equation}
{\bf F}\,\vec{u} = \lambda\, \vec{u}
\end{equation}
The eigenvectors can be used to diagonalize the Fisher matrix according to the following property 
\begin{equation}
\bf F = R^T\,\Lambda\,R
\label{eq.diagonalise}
\end{equation}
where the columns of the matrix ${\bf R}$ are the (normalised) eigenvectors of ${\bf F}$ and 
$\bf \Lambda$ is a diagonal matrix with the eigenvalues in the diagonal.
Notice that ${\bf R}$ can be understood as a rotation matrix that can be used to express the vector
of our initial parameters, $\vec{\Theta}$, in the new diagonal basis $\vec{u}$,
\begin{equation}
\vec{\zeta} = {\bf R}\,\vec{\Theta}
\end{equation}
from where we obtain our new set of parameters in the diagonal basis, $\vec{\zeta}$. 

\subsection{Fisher matrix tomography}

To compute the Fisher matrix we need to follow Eq.(\ref{eq.Fisher}).  We notice though that even for this simplified problem the straightforward application of this expression leads to long expression that make difficult a further analytical treatment.
To avoid cumbersome expression as much as possible we expand the Fisher matrix in its different composing terms. In the particular case of an experiment with M inputs and N outputs, we may write the elements of our Fisher matrix as:
%\begin{eqnarray}
%F_{ij} & = & \frac{1}{\Sigma_{11}} \left( \partial_{\theta_i} H_{11}(\theta) \right)^T \partial_{\theta_j} H_{11}(\theta) \nonumber \\
%& + & \frac{\Sigma_{12} +\Sigma_{21} }{\Sigma_{12} \,\Sigma_{21} } \left( \partial_{\theta_i} H_{21}(\theta) \right)^T \partial_{\theta_j} H_{21}(\theta) \nonumber\\
%& + & \frac{\Sigma_{12} +\Sigma_{21} }{\Sigma_{12} \,\Sigma_{21} } \left( \partial_{\theta_i} H_{12}(\theta) \right)^T \partial_{\theta_j} H_{21}(\theta) \nonumber\\
%& + & \frac{1}{\Sigma_{22}} \left( \partial_{\theta_i} H_{22} (\theta) \right)^T \partial_{\theta_j} H_{22} (\theta)
%\end{eqnarray}
\begin{equation}
F_{ij} =  \sum^{M}_{n,q =1}  \sum^{N}_{m,p =1}  F_{mnpq,ij} 
\end{equation}
where
\begin{equation}
F_{mnpq,ij}  =  \lbrace  {\bf \Sigma}^{-1} \rbrace_{mp}
\left[ \partial_{\theta_i} H_{mn}(\Theta) \right]^T \left[  \partial_{\theta_j} H_{pq}(\Theta)\right]
\,s_{n}\,s_{q}
\label{eq.termDef}
\end{equation}

The definition of the Fisher matrix allows us to combine the information of different experiments by adding their Fisher matrices. However, in this case, we use this same property in the opposite direction: to split a single experiment as the combination of simpler independent experiments.
This tomography will be particularly useful to interpret the Fisher matrix since we can split each experiment into the contribution of each transfer function and study them independently.
% It will also allow us to study different configurations of the same experiment and to proceed sequentially from 1D experiments to more complicated situations.
The $F_{mnpq,ij}$ term can be understood as the $mp-$component of a Fisher matrix corresponding
to an experiment which only considers a sinusoidal input applied to the  $nq-$channels. 
We notice here that if the noise covariance matrix, $\lbrace {\bf \Sigma}^{-1} \rbrace_{mp}$,
would be diagonal we could consider each $F_{mnpn,ij}$ as the contribution corresponding
to a given transfer function $H_{mn}(\Theta)$. However, cross-couplings between 
our channels imply a mixing of the different transfer function contributions.

\section{Design of input signals}
\label{sec.design}

The experiment design problem can be stated as how to choose an input signal 
that allows the optimisation of a given figure of merit, provided some constraints of our particular experiment. 
In literature~\cite{Mehra74}, there are several options for a scalar figure of merit to use as a minimisation criteria including i) the minimisation of the trace of the covariance matrix (A-optimality), thus minimising the average variance of the parameters; ii) the minimisation of the largest eigenvalue of the covariance matrix (E-optimality), which implies minimising the major axis of the uncertainty ellipsoid in the parameter space or iii) minimising the determinant of the covariance matrix (D-optimality), which is the equivalent to minimising the uncertainty ellipsoid in the parameter space. In the following, we will stick to the latter criterion since, among other advantages, it remains invariant under scale changes in the parameters. 

An analytical solution to the problem, as the one proposed in the previous section, has a limited application and becomes unfeasible for complex systems. The usual strategy is to describe the problem as a numerical minimisation problem as we show below.
For computational simplicity, the inverse of the Fisher matrix is used as an approximation of the covariance matrix. Since we are working in a high SNR regime, it is also a good approximation.

%A more flexible and general approach is to iteratively optimise the input spectrum based on a scalar criterion.
% As we discussed above, a useful choice for this purpose is to use the determinant of the covariance matrix which is a measure of the volume of the uncertainty ellipsoid. 
% This can be done analytically for simple models although no closed form exist for most problems. 

% \subsection{Dispersion function}

For mathematical convenience, our description of the system under study will be in frequency domain. Hence, recalling Eq.~(\ref{eq.system}), the input to our system will be described as 
\begin{equation}
\chi(\omega) = (|s(\omega_1)|^2 \ldots |s(\omega_F)|^2 ) 
\end{equation}
with bounded energy,
\begin{equation}
\sum^F_{k=1} |s(\omega_k)|^2 = 1
\end{equation}
where  $s(\omega_k)$ is the frequency domain representation of a given input at frequency $\omega_k$. Following the literature, we will call $\chi(\omega)$ our design. 
Different conditions can be set on the design in order to achieve D-optimality. 
Indeed, it can be shown that a design maximising the determinant of the Fisher 
matrix will minimise the maximum of the quantity~\cite{Goodwin}
\begin{equation}
\nu(\omega) = {\rm tr} \left[ { \bf F^{-1}(\chi) F(\omega) }\right]
\label{eq.dispersion}
\end{equation}
where $ \bf F(\chi)$ is the information matrix from the design $\chi(\omega)$
and $ \bf F(\omega)$ is the information matrix from a single frequency 
input with normalised power spectrum $|X(\omega)|^2 = 1$. 
The quantity $\nu(\omega)$, known as dispersion function or  response dispersion, 
can be understood as the ratio of the 
variance of the system transfer function to the noise power.

Based on its mathematical properties, the dispersion function
has been proposed as a tool for input design optimisation. The underline
idea is to select a frequency grid where the power of the input signal is 
initially uniformally distributed among the the selected frequencies. 
The dispersion function is then computed for each frequency in the grid 
and the power of the signal distributed proportionally to the value of this 
function. The optimal design is achieved by repeating this procedure iteratively.
More precisely, the algorithm steps are~\cite{WalterPronzato, SchoukensPintelon}:
\begin{enumerate}
\item Select a set of frequencies $\lbrace \omega_1, \ldots \omega_F \rbrace$ 
within the frequency band of interest and 
distribute the power equally over these frequencies. This constitutes the initial design.
\item Compute the dispersion function for the F frequencies.
\item Create a new design according to:
\begin{equation}
\chi_{i+1}(\omega_k) = \chi_{i}(\omega_k) \times  \nu_i(\omega_k)/\rm N_{\theta}
\end{equation}                                                                                                                                                                                                                                                                                            
\item If $\max(\nu(\chi_i,\omega_k)) - N_{\theta}) <\epsilon$ for a sufficient small $\epsilon$, 
then the optimum design is found. 
If not, we return to step 2.
\end{enumerate}

It can be shown that the algorithm converges to a D-optimal design.~\cite{SchoukensPintelon}.
                                                                                                                                                                                                                                                                                             
% \subsection{An example: the Harmonic Oscilator ~\label{sec.harmOsc}}

In order to prove the efficiency of the previous numerical design 
method we test it in the case of a harmonic oscillator.
We can analytically compute the Fisher matrix for this problem to obtain an expression 
which, as expected, shows a maximum of the spectrum at the natural frequency of the oscillator, $ \omega_0$. This value is therefore the one that minimises the volume of the error ellipsoid in the 
parameter space and hence, the one that the numerical method described in the 
previous section should retrieve.
%With that end we consider a system where we describe
% the output $y$ in frequency domain as
%\begin{equation}
%y =  \frac{\omega^2_0}{-\omega^2 + 2 \imath \xi \omega_0\, \omega +\omega^2_0} \, x + n
%\end{equation}
%where $x$ is the injected input signal and n is white noise with variance $\sigma$. In the
%system transfer function, we only consider the damping ration, $\xi$, and the 
%natural frequency of the oscillator,$\omega_0$, as parameters of 
%our problem. 

\begin{figure}[t]
\begin{center}
\includegraphics[width=0.49\textwidth]{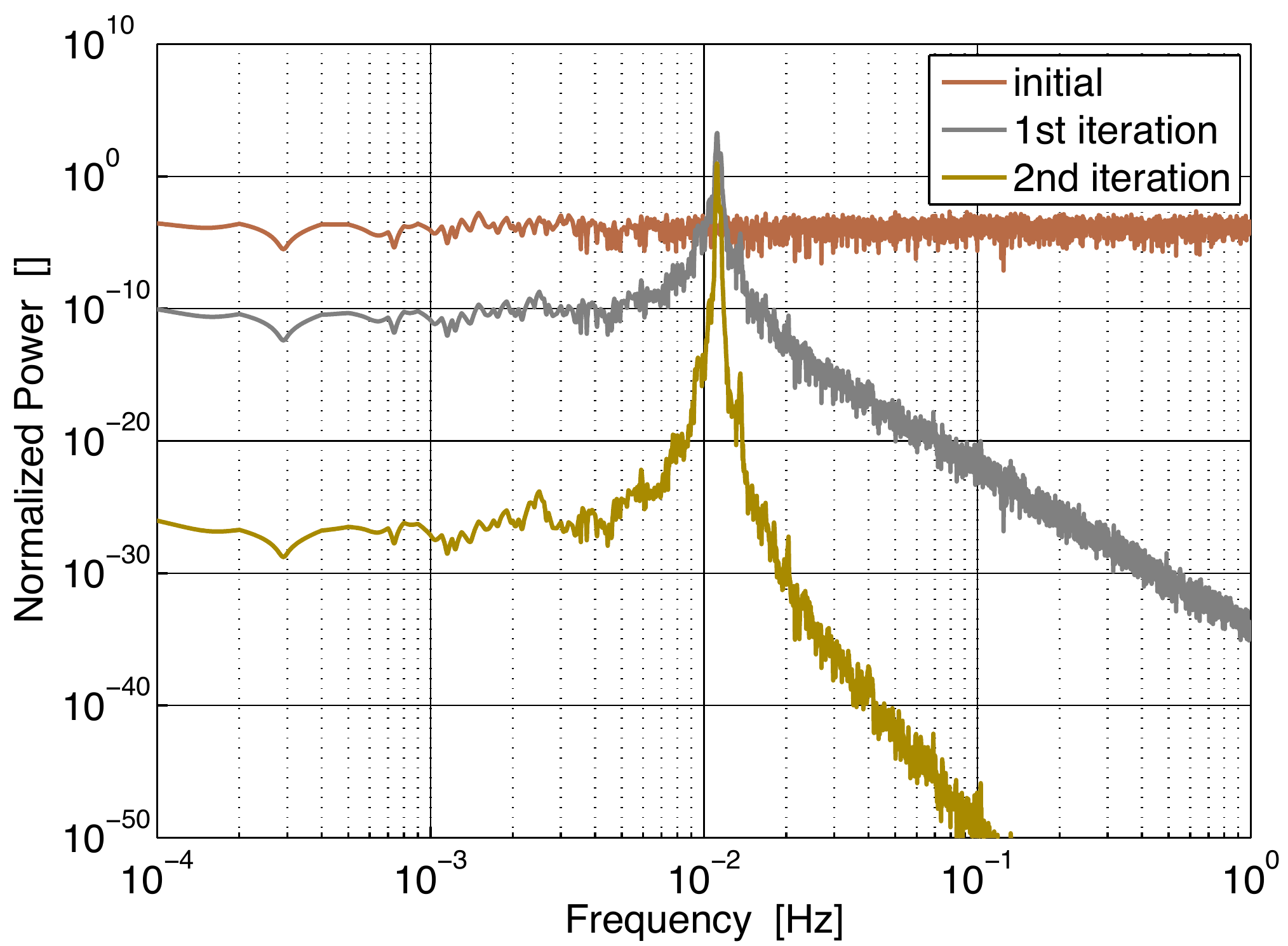} 
\caption{Evolution of the algorithm to optimise the input signal for the harmonic oscillator case. The algorithm promotes the natural frequency of the oscillator $\omega_0 = 0.07$. \label{fig.harmosc}}
\end{center}
\end{figure} 

%\begin{equation}
%\det \left[ {\bf F} \right] = \frac{1}{\sigma}\frac{16\, \omega ^6 \omega_0 ^8}{\left(\omega ^4+2 \left(2 \xi ^2 - 1\right) \omega ^2 \omega_0^2+\omega_0^4\right)^4}
%\end{equation} 

In order to check the validity of our methodology, we generated a time series of 10\,000 seconds
of white noise with variance $\rm \sigma = 10^{-5}$ that we consider as our initial input design. 
We choose white noise in order to weight all frequencies equally. We consider 
an harmonic oscillator with damping ratio $\xi = 0.01$ and natural frequency $\omega_0 = 0.07$, and then we run the algorithm as described above. The result is shown in Fig.~\ref{fig.harmosc} where we show the evolution of the input signal as
proposed by the algorithm. As shown, two iterations are enough for the algorithm 
to promote the natural frequency of the oscillator $\omega_0$ among the others.

\section{LISA PAthfinder model \label{sec.LPFmodel}}

In order to apply this methodology to LISA Pathfinder we will need first to
define a model for the experiment. In the following we introduce the notation
to describe the combined dynamics of the two test masses and the satellite required for 
the analysis.   
The same description with small variations can also be found in~\cite{Nofrarias10,Ferraioli11,Congedo12}.
\subsection{Equation of motion \label{sec.eqmotion}}
The measurement on-board the satellite is usually expressed as
\begin{eqnarray}
  \vec{o}  & = &  ({\mathbf{D}}\cdot{\mathbf{S}}^{-1}+{\mathbf{C}})^{-1} (-
{\mathbf{C}}\,\vec{o}_i + \vec{g}_n + {\mathbf{D}}\cdot {\mathbf{S}}^{-1} \vec{o}_n)
\label{eq.dyn}
\end{eqnarray}
where $\mathbf{D}$ is the dynamical matrix, $\mathbf{C}$ is the
controller, and $\mathbf{S}$ stands for the sensing matrix, 
which translates the physical position of the test masses into
the interferometer readout, $\vec{o}$. Subindex~$n$ stands for noise
quantities, either sensing noise ($\vec o_n$) or force noise ($\vec
g_n$), and subindex~$i$ stands for the injected signals ($\vec
o_i$). 
Restricting ourselves to linear motion along the axis between the two test masses (the degree of freedom that is measured by the interferometer), each of the dynamical variables in (\ref{eq.dyn}) can be expressed as 2-dimensional vectors with components referring to 
the  $\mathrm{x}_1$ and $\mathrm{x}_{\Delta}$ channels respectively,
\begin{eqnarray}
    \vec{o}  = \left(
  \begin{array}{c}
   o_{1} \\ 
    o_{\Delta} 
  \end{array}   \right), \nonumber \quad &  &
  \vec{o}_i  = \left(
  \begin{array}{c}
   o_{i1} \\ 
   o_{i\Delta}  
  \end{array}   \right), \\
  \vec{o}_n  = \left(
  \begin{array}{c}
   o_{n1} \\ 
   o_{n\Delta}  
  \end{array}   \right), \nonumber \quad &  &
    \vec{g}_n  = \left(
  \begin{array}{c}
   g_{n1} - g_N \\ 
   g_{n2} - g_{n1}  
  \end{array}   \right), 
  \label{eq.sigVec}
\end{eqnarray}

where subindicies 1 and 2 refer to the first and second test mass, subindex $\Delta$ refers to differences between the first and second test mass, and capitalised subindices (such as force noise on the spacecraft, $g_N$) refer to the spacecraft. The last equation in (\ref{eq.sigVec}) shows how $g_N$ is only measured in the first channel. On the other hand,
the differential channel is sensitive to any differential force noises applied to 
the first and the second test masses.

%In the simplified one dimensional LTP model used in the following we consider 2 possible injection inputs 
%and 2 measurement channels. The parameter set considered is the following $\Theta~=~\lbrace G_{\rm df}, \omega^2_1, G_{\rm lfs}, \omega^2_2, \delta_{21} \rbrace$

The matrices describing the dynamics of the LISA Pathfinder system are:
\begin{eqnarray}
  \mathbf{D}  & = & \left(
  \begin{array}{cc}
  s^2 + \omega^2_{1}  + \frac{m_1}{m_{\rm SC}} \omega^2_1 + \frac{m_2}{m_{\rm SC}} \omega^2_2 & \frac{m_2}{m_{\rm SC}} \omega^2_2 \\ 
  \omega^2_{2} -\omega^2_{1} &  s^2 + \omega^2_{2}
  \end{array}   \right), \nonumber \\ \nonumber \\
  \mathbf{C} & = & \left(
  \begin{array}{cc}
   H_{\rm df} & 0\\
   0 &  H_{\rm sus}
  \end{array}
  \right),  \label{eq.dyn} \\ \nonumber \\
  \mathbf{S} & = & \left(
  \begin{array}{cc}
  S_{11} & S_{12}\\
   S_{21} & S_{22}
  \end{array}
  \right), \nonumber
\end{eqnarray}
%\end{widetext}
where $\omega_{1}$ and $\omega_{2}$ are the stiffnesses ---
the steady force gradient across the test mass housing 
per unit mass~\cite{Bortoluzzi04}  ---
coupling the motion of each test mass to the motion of the spacecraft;
$H_{\rm df}$ and $H_{\rm sus}$ are the drag-free and suspension loops
controllers, respectively. For the remainder of this work, it is assumed that $H_{\rm df}$ and $H_{\rm sus}$ are known.

For our current analysis we will assume some approximations
in these expressions in order to keep the main scientific information
and, at the same time, keep the expressions as simple as possible.
For that reason, in the following we will eliminate the back reactions terms,
$m_1 = m_2 << m_{\rm SC}$,
consider that the sensing matrix cross-couplings are zero 
$S_{12} = S_{21} = 0$.
For convenience, we will take the calibrations $S_{11} = S_{22} = 1$.  
%\subsection{Transfer functions from input guidance}
Taking into account these assumptions we 
can derive expressions for the transfer functions describing the system. 
We consider input signals injected at the guidance
input port which we expressed as $o_i$ in Eq.(\ref{eq.dyn}), hence
the transfer function is defined by 
\begin{eqnarray}
{\bf H} & = &  ({\mathbf{D}}\cdot{\mathbf{S}}^{-1}+{\mathbf{C}})^{-1} (-\mathbf{C})\cdot\vec{o}_i \nonumber \\
& = & 
\left(
\begin{array}{cc}
H_{11}(\Theta) & H_{12}(\Theta)  \\
H_{21}(\Theta)  & H_{22}(\Theta) \\
\end{array}
\right)
\left(
\begin{array}{c}
o_{i1} \\
o_{i\Delta}
\end{array}
\right)
\label{eq.trnsfer}
\end{eqnarray}
where the transfer functions are given by
\begin{eqnarray}
H_{11} &=& \frac{H_{\rm df}}{\omega^2 - \omega^2_1 + H_{\rm df}}\\ 
H_{12} &=& 0\\ 
H_{21}  &=& \frac{H_{\rm df} \, (\omega^2_2 - \omega^2_1)}{(\omega^2 - \omega^2_1 + H_{\rm df})(\omega^2 - \omega^2_2 +H_{\rm lfs})}  \label{eq.H21} \\
H_{22}&=&\frac{H_{\rm lfs}}{\omega^2 - \omega^2_2 + H_{\rm lfs}} \label{eq.H22}
\end{eqnarray}
where we realise that $H_{12}$ is zero because this is proportional to the parameter $S_{12}$, 
which is considered to be zero. At the same time, we see that the cross-coupling from drag-free to 
differential channel, $H_{21}$, is proportional to the differential stiffness, $\omega^2_2 - \omega^2_1$.

\subsection{Noise model}
Our study of the injection scheme in LISA Pathfinder relies on the Fisher matrix 
which, in turn, depends on the noise model used for those noise sources identified in Eq.(\ref{eq.dyn}).
These are: 
interferometer read-out noise for both channels ---$o_{n1}$ and $o_{n\Delta}$--- 
force noise applied to the test masses ---$g_{n1}$ and $g_{n2}$--- and 
force noise applied to the spacecraft ---$g_{N}$. 
Following~\cite{Nofrarias10}, we will characterise each of these 
with the five parameters, $p_{1\dots~5}$ in the expression
\begin{equation}
S(\omega) = p_1 \, 
\left( 1 + \frac{1}{(\frac{\omega}{2 \pi \, p_2 })^{p_3}}
+ \frac{1}{(\frac{\omega}{2 \pi \, p_4 })^{p_5}}\, 
\right)^{1/2}.
\end{equation}

\begin{figure}[t]
\begin{center}
\includegraphics[width=0.49\textwidth]{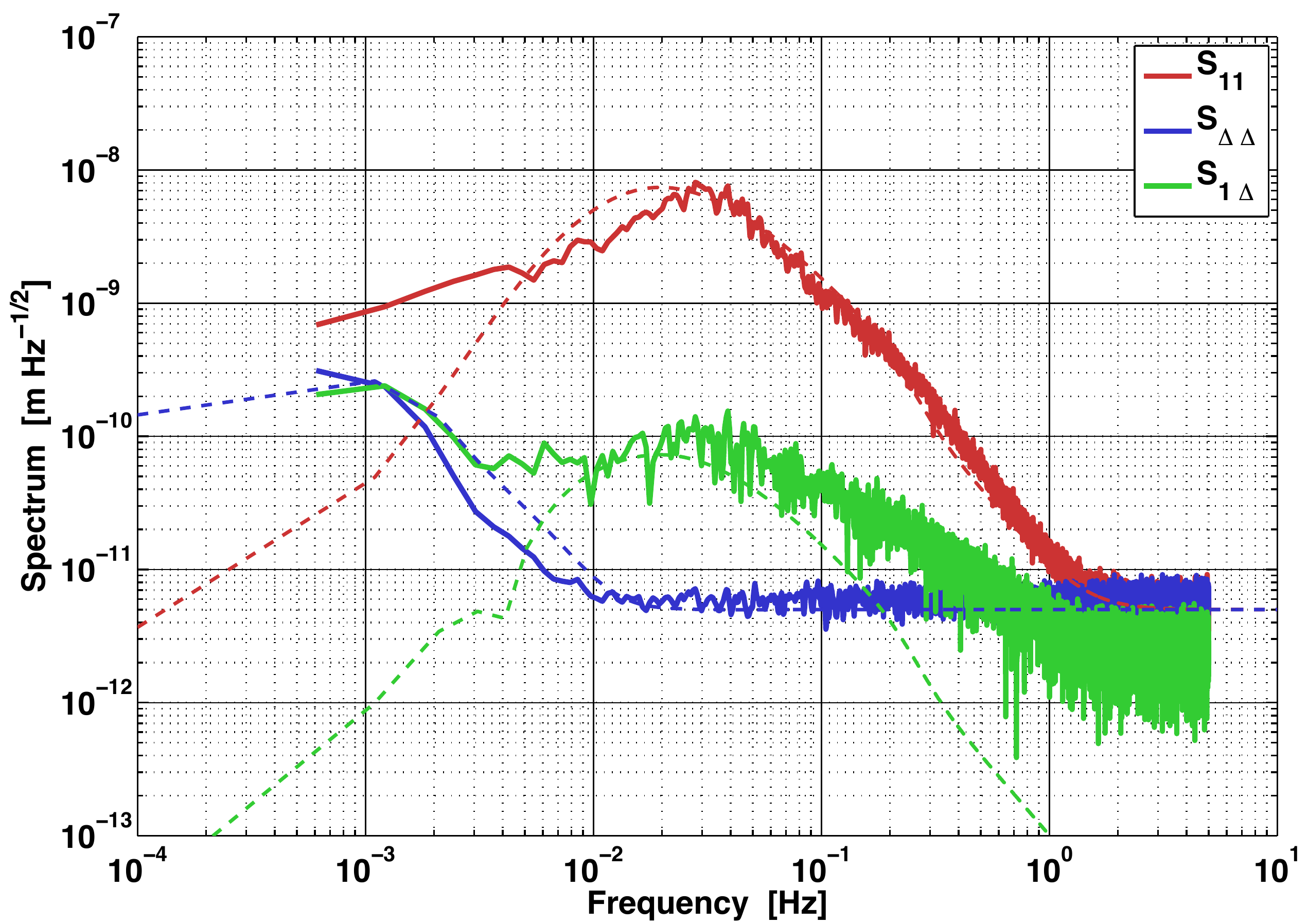} 
\caption{Comparison of the noise spectra for the two main interferometers channels 
for an analytical simple model (dashed line) and a noise data stream generated via a LPF state-space model (solid line). 
 \label{fig.noise}}
\end{center}
\end{figure}

Applying the parameters in Table \ref{tbl.model} we obtain the models in Fig.~\ref{fig.noise} for the noise spectra of the two main interferometer channels.  
We can compare the predictions from this simplified model
to simulations coming from a detailed state-space
simulator containing a much elaborate model of the instrument, for
instance delays, actuators, and component noise models~\cite{Nofrarias13}.
As seen in Fig.~\ref{fig.noise}, our simple parametric model agrees well with the noise obtained from the state-space model.

\begin{table}[t]
\caption{LPF noise model parameters \label{tbl.model}. $p_2$ and $p_4$ parameters correspond to 
frequencies in [Hz] and $p_1$ to amplitude spectral densities in [$\rm m/\sqrt{Hz}$] and [$\rm N/\sqrt{Hz}$] for
read-out noise and force noise, respectively.}
\begin{ruledtabular}
\begin{tabular}{lccc}
 \multicolumn{4}{c}{\textsc{Noise Parameters}}\\
\hline
Parameter & $o_{n1}/o_{n \Delta}$ & $g_{n1}/g_{n2}$ & $g_N$ \\
\hline
$p_1$ & $3.6\times 10^{-12}$  & $7\times 10^{-15} $ &  $2.5\times 10^{-10}$  \\
$p_2$ & $10\times 10^{-3}$     & $ 5\times 10^{-3}$   &  $12\times 10^{-3}$ \\
$p_3$  & 4.2                             & 3                              &  3.8 \\
$p_4$  & $1.8\times 10^{-3}$   & $4\times 10^{-4}$    &  $1\times 10^{-3}$\\
$p_5$ & 8                                 &  8                             & 8 \\
\end{tabular}
\end{ruledtabular}
\end{table}

\section{Calibration signals for LISA Pathfinder dynamics \label{sec.LPFOptSig}}

During operations, LISA Pathfinder will run an exhaustive characterisation campaign 
with the objective of calibrating the instrument and identifying the main noise contributions.
Here we consider one set of experiments targeting the calibration 
of the dynamical parameters governing the 
combined motion of the two test masses and the satellite. 
For these particular set of experiments, the calibration procedure consists of the injection of a 
sequence of sinusoids --- the only available waveform in the flight software --- at different frequencies at a number of input ports. For this work, we will focus on injection in one of the two main interferometer 
channels. However, the methodology can be easily applied to the remaining degrees of freedom.

\begin{figure*}[t]
\begin{center}
\includegraphics[width=0.49\textwidth]{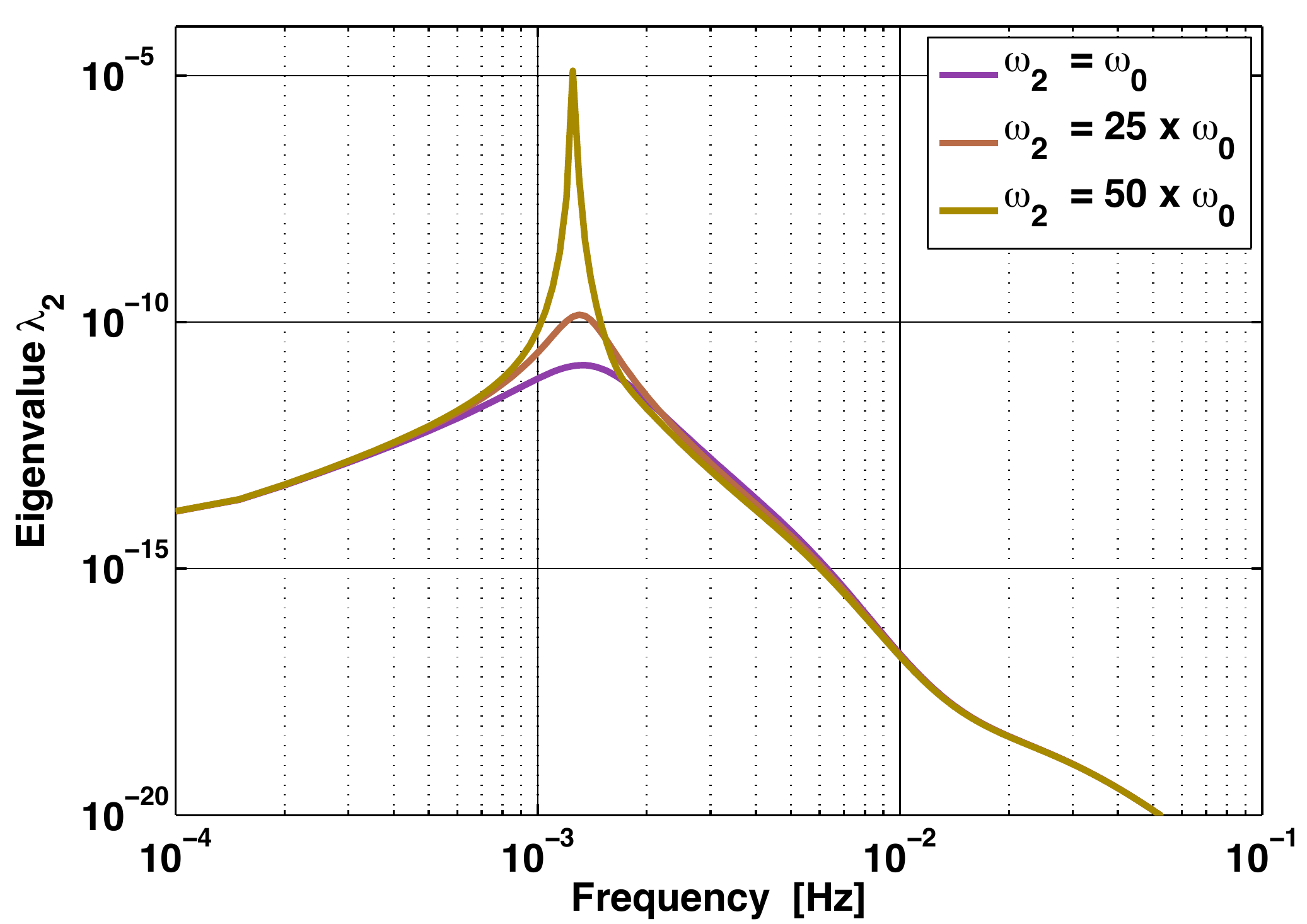} 
\includegraphics[width=0.49\textwidth]{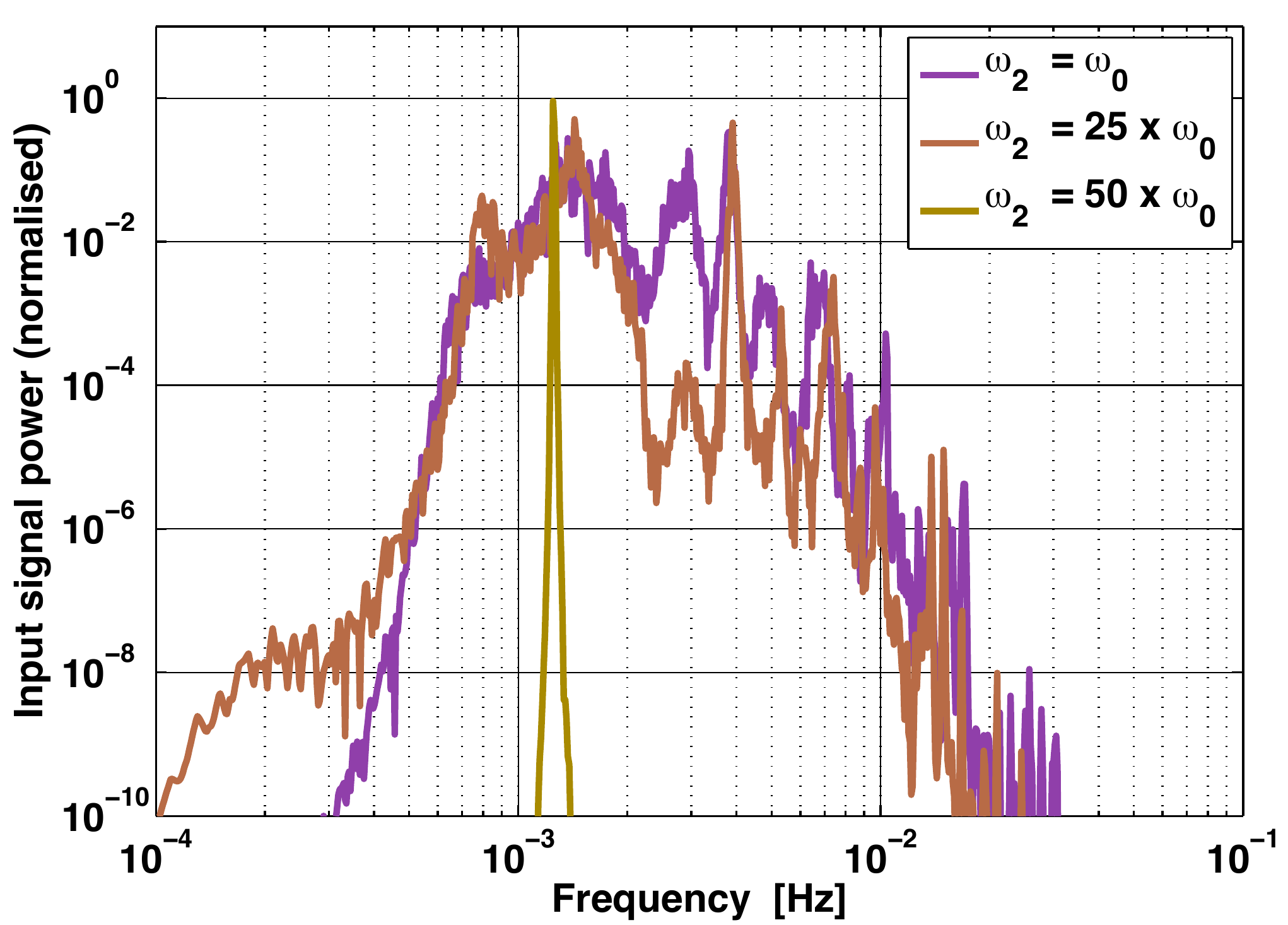}
\caption{\emph{Left}: evaluation of the eigenvalue $\lambda_2$ of the $F_{2121}$ term for a single frequency injection. \emph{Right}: Output of the numerical algorithm for the optimisation of input signals based on the dispersion function applied to a LISA Pathfinder state-space model after 25 iterations. In this case, white noise was injected into the drag-free channel. In both cases (analytical and numerical) the analysis is repeated by rescaling a factor 25 and 50 the value of the second test mass stiffness, which is originally considered to be $\omega^2_0 = -22\times 10^{-6}\,\rm s^{-2}$.  \label{fig.lambda2}}
\end{center}
\end{figure*}

\subsection{The~$F_{2121}$ term }
In order to demonstrate our method we consider an injection applied to the drag-free channel. In our framework this experiment would be completely described by the sum
\begin{equation} 
 F_{ij}    =  \sum_{m,p} F_{m1p1,ij}  \label{eq.Fij}
\end{equation}
where the indices $i$ and $j$ run over the parameters. The most general case (7 degrees of freedom) correspond to 49 terms. This is not approachable analytically so we focus our attention on one term with particular relevance, the $ F_{2121}$, which can be expressed as:
\begin{eqnarray}
F_{2121,ij}  & = &   \lbrace {\bf \Sigma}^{-1} \rbrace_{22} \times \nonumber \\
& & \left[ \partial_{\theta_i} H_{21}(\omega) \right]^T 
\left[  \partial_{\theta_j} H_{21}(\omega) \right]
\,| o_{1}(\omega) |^2
\label{eq.F2121}
\end{eqnarray}
This term quantifies the effect of the injection in the first channel as measured by the highly-sensitive differential channel. Under the assumptions discussed in sec.~\ref{sec.eqmotion}, the only parameters that impact this term are the two test mass stiffnesses, which enter through the term in Eq.~(\ref{eq.H21}).
%\begin{equation}
%G_{21}  = - \frac{H_{\rm df}(\omega)\, (\omega^2_1 - \omega^2_2)}{(H_{\rm df}(\omega) - \omega_1^2 + \omega^2)\,(H_{\rm lfs}(\omega) - \omega_2^2 + \omega^2)}
%\end{equation}
Due to this simplification, we can describe this problem 
in analytical terms. 
%The covariance term $ {\bf \Sigma}^{-1} \rbrace_{22} $ 
%can be found in Appendix \ref{sec.AppNoise}
Eq.(\ref{eq.F2121}) turns into a $2\times 2$  
matrix that we can easily decompose in the related eigenvectors,
\begin{equation}
\vec{u_1} = 
\left[ 
\begin{array}{c}
\dfrac{\alpha(\omega_1)\,\beta(\omega_1)}{\alpha(\omega_2)\,\beta(\omega_2)} \\
1 \\
\end{array} \right]
\quad
\vec{u_2} = 
\left[ 
\begin{array}{c}
-\dfrac{\alpha(\omega_2)\,\beta(\omega_2)}{\alpha(\omega_1)\,\beta(\omega_1)} \\
1
\end{array}
\right]
\end{equation}
where
\begin{eqnarray}
\alpha(x) & = & H_{\rm df}(\omega) - x^2 + \omega^2 \\
\beta(x) & = & H_{\rm lfs}(\omega) - x^2 + \omega^2
\end{eqnarray}
and associated eigenvalues,
\begin{eqnarray}
\lambda_1 & = & 0 \nonumber \\
\lambda_2 & = &  \lbrace {\bf \Sigma}^{-1} \rbrace_{22} \, H^2_{\rm df}(\omega)\, \frac{\alpha^2(\omega_1)\,\beta^2(\omega_1) \,\alpha^2(\omega_2)\, \beta^2(\omega_2)}{\alpha^4(\omega_1) \beta^4(\omega_2)} \label{eq.lambda2}
\end{eqnarray}
Since eigenvalues are directly related to the inverse 
of the expected uncertainty on the associated parameter, 
we conclude that this measurement can only constrain parameters in the $\vec{u}_2$ direction while the direction $\vec{u}_1$ has an associated uncertainty 
that tends to infinity.

It is important to notice at this point that the analysis we perform in the frequency 
domain implicitly assumes a unique frequency, i.e. 
an input signal which is a sinusoid at a given frequency. 
In the following section we explore which information
can we obtain in such a case. 
%Later, we expand the analysis
% to a sequence of sinusoids with different frequencies.

\subsubsection{Single tone input: undetermined solution \label{sec:inj2_1sin}}
If we assume that the content of our input signal is a sinusoid with a fixed frequency, 
we know from the previous eigendecomposition that we will not be able to 
solve the problem since we have only one valid eigenvector for a 2-dimensional 
problem. Nonetheless, we explore the single frequency solution in order
to determine how much information can we get from the system in such a case.

\begin{figure*}[t]
\begin{center}
\includegraphics[width=0.5\textwidth]{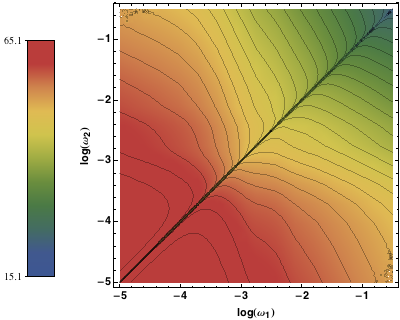} 
\includegraphics[width=0.49\textwidth]{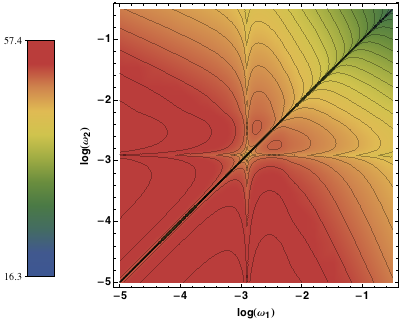} 
\caption{Determinant of the $F_{2121}$ term for an injected signal with two independent
frequencies. The determinant is evaluated for the case $\rm \omega^2_2 = -22 \times 10^{-7}\,s^{-2}$ (left) and $\rm \omega^2_2 = 50 \times (-22 \times 10^{-7})\,s^{-2}$ (right). The values on the legend are the          $\rm log \left( det \left[F_{2121} \right] \right)$.
 \label{fig.F2121det} }
\end{center}
\end{figure*}

We proceed to diagonalise the Fisher matrix as in Eq.~(\ref{eq.diagonalise}) 
\begin{equation}
\bf F = R^T\,\Lambda\,R \nonumber
\end{equation}
from which we obtain a diagonal system with a unique eigenvalue, $\lambda_2$, given by equation Eq.(\ref{eq.lambda2}). In Fig.~\ref{fig.F2121det} we explore this expression as a function of the frequency 
of the injection. We see that the eigenvalue has a peak when the input is injected at a frequency around f = 1.25\,mHz. This becomes more evident if we increase the value of $\omega^2_2$, as shown in the figure. 

The value of f = 1.25\,mHz is therefore the best frequency for a signal composed with a unique frequency component for the experiment under study. Indeed, by maximising the Fisher matrix we are reducing the error on the parameter space. However, it must be noted that this is not necessarily an optimal solution since we are dealing with a single frequency injection scheme that leads to singular Fisher matrix. 
%In section~\ref{sec:inj2_2sin} we move to the two frequencies injection, overcoming this constraint.

A second consideration to take into account is that when diagonalising our system, our parameters are expressed in a new basis which corresponds to applying a rotation matrix {\bf R} to the original vector of parameters $\vec{\Theta} = \lbrace \omega^2_1, \omega^2_2 \rbrace$. In doing so, we obtain a new set of parameters $\vec{\zeta} = {\bf R} \cdot \vec{\Theta}$. For the configuration under study the combination of parameters corresponding to the non-zero eigenvalue is proportional to the sum of stiffness, i.e.
\begin{equation}
\zeta_2 \propto \omega^2_1 + \omega^2_2
\end{equation} 
confirming that a single frequency signal is not able to break the degeneracy between the two parameters in our system.

We are now prepared to compare the results obtained analytically 
with the prediction of the numerical algorithm based on the dispersion function --- Eq.~(\ref{eq.dispersion}). To do so we inject 
a white noise data stream to the input channel under consideration, which for the 
analysis of the $F_{2121}$ term is the drag-free channel. In this particular case, 
we consider as  our initial input a white noise time series of $10^5$\,s and 
$\rm \sigma = 10^{-6}\,m^2$. 

In the right panel of Fig.~\ref{fig.lambda2} we show the resulting normalised power spectrum 
of the input signal as retrieved after 25 iterations of the numerical optimisation algorithm.
The algorithm promotes the same frequencies that maximised the eigenvalue of the Fisher matrix 
$F_{2121}$ as can be seen in the left hand figure. Moreover, we performed the analysis by rescaling the $\omega^2_2$ value as in the study of the eigenvalues. Here we observe again how the numerical algorithm selects the f = 1.25\,mHz frequency when approaching the case where $\omega^2_2$ is rescaled by a factor of 50, proving the consistency between the analytical and the numerical approach.

It is worth stressing the agreement between the two approaches shown, given that they are not based in the same description of the instrument. Whilst the analytical derivation is funded in the expressions derived here from Eq.~(\ref{eq.dyn}), the numerical approach has its roots in the numerical computation of the dispersion function Eq.~(\ref{eq.dispersion}) which uses a state space representation of LISA Pathfinder. 
The difference also lies oi how the instrument noise enter in the analysis. While we analytically compute the term ${\bf \Sigma}_{22}$ in Eq.(\ref{eq.lambda2}), the noise enters in the numerical analysis through the evaluation of the Fisher matrix in Eq.~(\ref{eq.dispersion}). In the later, the noise spectrum of the instrument is computed by generating time series with the LPF state-space model configured with no signal injections and then computing the power spectrum.

%These would represent the situation where the two frequencies
%of the injected sinusoid would be equal. As shown in section~\ref{sec:inj2_1sin}, a single sinusoid 
%injection yields a Fisher matrix with rank~=~1 which explains the sudden drop of the determinant 
%since the determinant of a non-full-rank determinant is exactly zero. This is observed as well in the 
%\emph{valley} on the diagonal of both panels in Fig.~\ref{fig.F2222det}.

\subsubsection{Two tones input: full-rank solution \label{sec:inj2_2sin}}

\begin{figure*}[t]
\begin{center}
\includegraphics[width=0.8\textwidth]{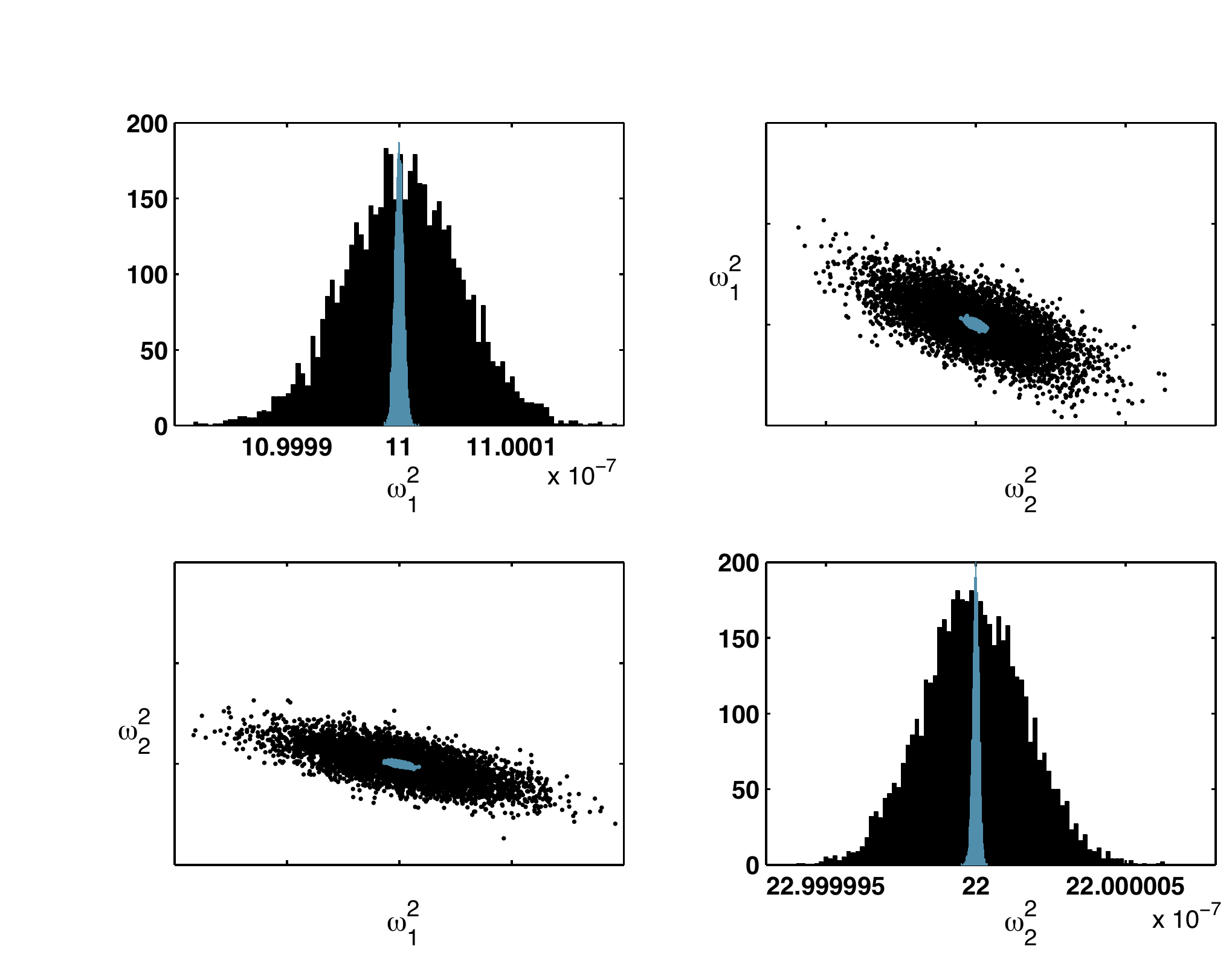} 
\caption{Expected error on parameters for an injection in the drag-free channel considering  $\omega^2_1$ and $\omega^2_2$ as the only relevant parameters. Black corresponds to the initial proposal of a white noise input, blue represents the expected error for the input signal as obtained with the proposed numerical algorithm after 25 iterations. Histograms are computed based on 5000 samples of a multivaritate gaussian distribution with the covariance matrix obtained by the numerical algorithm.  \label{fig.covMatrix}}
\end{center}
\end{figure*}

Here we take advantage of the analytical solution to go one step further and explore the
case of an input signal composed by two sinusoids. 
In order to combine the information of more than one sinusoid frequency in the 
input signal we add the Fisher matrices corresponding to each frequency. 
Our experiment will therefore be described by
\begin{equation}
F_{2121,ij} = \sum^{N=2}_{k} \bar F_{2121,ij}(\omega_k)
\label{fig.2sinF2121}
\end{equation}
where each $\bar F_{2121,ij}(\omega_k)$ corresponds to the contribution 
of a single sinusoid injection to the final experiment's Fisher matrix. 

We first explore the rank of the $F_{2121}$ matrix when evaluated for different
combinations of these two input frequencies. Given that $F_{2121}$ depends on two parameters,
results show that most combinations of frequencies are able to reach the condition 
${\rm rank}(F_{2121}) \ne 2$. In fact, only when the two frequencies are equal 
--and we come back to our previous case-- we will not be in a full-rank situation.
This allows us to go one step further and explore which combination of frequencies
are optimal, in the sense of maximising the Fisher matrix, i.e. minimising the ellipsoid 
error volume in the parameters space.  
Figure \ref{fig.F2121det} shows the value for the determinant of the $F_{2121}$ term as a function 
of the two injection frequencies. 
We explore the determinant for two different configurations of the experiment: the standard with 
$\rm \omega^2_2 = -22 \times 10^{-7}\,s^{-2}$ and, as before, rescaling  $\rm \omega^2_2 = 50 \times (-22 \times 10^{-7})\,s^{-2}$. As expected, the determinant shows symmetry since the two injection frequencies in Eq.(\ref{fig.2sinF2121}) can be interchanged producing the same output. The determinant drops to zero at the diagonal since, as commented above, an injection with two equal frequencies sinusoid does not lead to a full rank solution.
It is interesting to see that when we set $50 \times \omega^2_2$, a notch appears at the frequency f = 1.25\,mHz that we found as a maximum in the single injection case.  

In the standard configuration, the maximum of the $F_{2121}$ determinant appears for frequencies in the very low frequency regime ($\rm f < 1$\,mHz). If, for practical reasons, we set one of the two injections to be $\rm f_1 = 0.1$\,mHz the maximum of the function displayed in Figure \ref{fig.F2121det} appears for a second injection at $\rm f_2 = 0.3$\,mHz.
With these two values we can proceed to estimate the expected errors on the parameters, by evaluating the Fisher matrix in Eq.(\ref{fig.2sinF2121}). 
By assuming two sinusoid injections with two cycles each at the obtained frequencies $\rm f_1 = 0.1$\,mHz and $\rm f_1 = 0.3$\,mHz with and amplitude of $10^{-7}$\,m, we can evaluate our expression for the Fisher matrix term $F_{2121}$, obtaining a $7 \times 10^{-3}\,\%$ and $6 \times 10^{-3}\,\%$ relative error estimate for the two stiffness parameters $\omega^2_1$ and $\omega^2_2$, respectively. It is worth reminding here that these are optimal errors representing the contribution of the $F_{2121}$ term of the Fisher matrix to the overall experiment. We consider it as a useful example to show the capability of the framework here proposed to disentangle the different contributions to the experiment. However, the precise determination of the  expected error for a given parameter requires the evaluation of the full Fisher matrix, which is composed in the analytical description of 49 components for the drag-free injection experiment. Hence, analysis considering the whole system are, in most cases, more suited for a numerical approach.

In order to evaluate the improvement on the estimate of the parameters, we run the analysis using the numerical algorithm introduced in Sec.~\ref{sec.design} assuming an injection in the drag-free channel and considering only the two stiffness $\omega^2_1$ and $\omega^2_2$ as relevant parameters. As described above, the algorithm evaluates the Fisher matrix at each step so we can trace how the expected errors for each parameter improve by modifying the input signal. The improvement in the error, as given by the Fisher matrix, is shown in Fig.~\ref{fig.covMatrix}, where we compare expected error on the parameters at the 1st and at the 25th iteration. The input signals associated with these two cases corresponds to a white noise injection for the first iteration that turns into a signal focusing all the power at $\rm f = 1.25$\,mHz after 25 iterations. The results show a clear improvement in the expected error on the parameters which decreases roughly by an order of magnitude. 

\section{Robustness analysis}
\label{sec.robusteness}
A last consideration that has to be taken into account in the 
design of experiments is that of the robustness of the analysis.
A key issue of the experiment design framework is that 
it relies on the evaluation of the Fisher matrix, which depends 
on the true values of the system under study, precisely the 
unknowns that the experiment aims to identify. 
In most cases, some reasonable estimates for the expected values 
exist and therefore the experiment is designed on the basis of this
\emph{a-priori} knowledge. 
The assumption is thus that the design obtained will not show a strong 
dependence on the values being considered. 
However, this raises the question if the experiment 
being defined in such a way is merely reinforcing our
previous knowledge about the system~\cite{Goodwin06}.
This difficulty has been recognised in the literature 
and several approaches have been proposed in order
to achieve a robust optimal design scheme~\cite{Rojas07}, 
although methods that are robust with respect uncertainties in the
system parameters is a wide open research field~\cite{Hjalmarsson05}.

In the particular case of LISA Pathfinder, we focus 
on the sequential design scheme~\citep{Wu85, Ford85, WalterPronzato}
which proposes to overcome the circular reasoning above 
by iteratively switching from parameter estimation to experiment design 
using the most recent parameter estimates from the previous step 
to define the next experiment. 
Indeed, this experiment design strategy fits particularly for our mission 
scenario since it includes in a natural way the process of improving the 
knowledge of our system parameters that will occur during the mission. 
Moreover, the sequential design can be already tested against the mission 
simulators that we have previously introduced, as we show in the following. 

Let's consider a LISA Pathfinder state-space model where, 
instead of the two degrees of freedom considered before, 
we increase now the complexity 
of the problem to include five unknown parameters in the system, 
corresponding to the parameter set 
$\lbrace G_{\rm df}, G_{\rm lfs}, \omega^2_{1}, \omega^2_{2}, S_{21} \rbrace $
in Eq.~(\ref{eq.dyn}). 
That way, we allow our system to explore a wider parameter space. 

In order to quantify the robustness of our experiment design strategy 
we run a Monte Carlo analysis allowing the parameters of our model
to be uniformly distributed as 
\begin{eqnarray}
G_{\rm df} & \sim & G_{\rm lfs} \sim \mathcal{U}[0.9,0.2], \nonumber\\
\omega^2_{1} & \sim & \omega^2_{1} \sim \mathcal{U}[1.9 \times 10^{-6} , 1.1\times 10^{-6} ], \\
S_{21}  & \sim &  \mathcal{U}[1.55 \times 10^{-4} , 1.45\times 10^{-4} ]. \nonumber 
\end{eqnarray}

\begin{figure}[t]
\begin{center}
\includegraphics[width=0.5\textwidth]{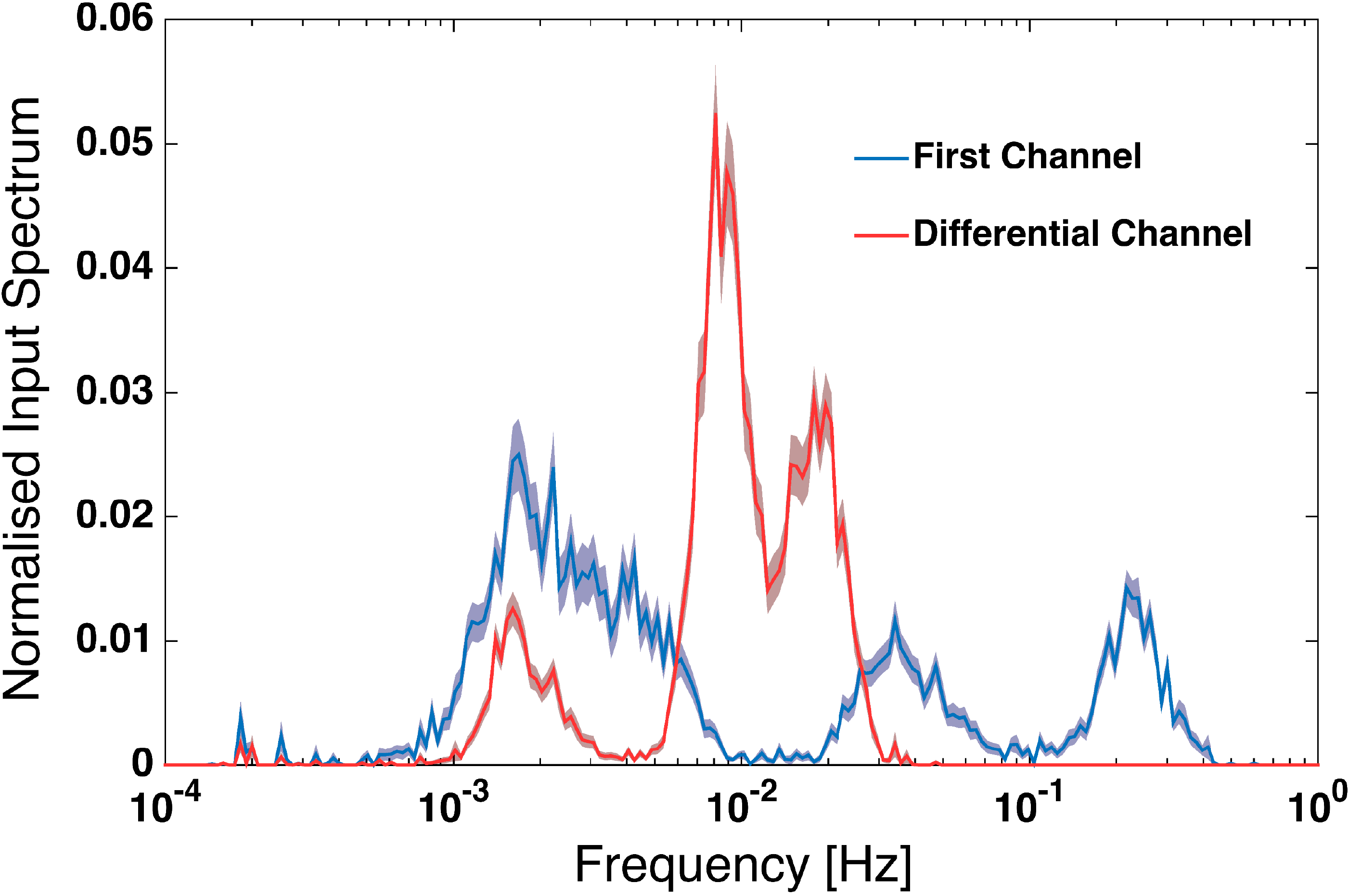} 
\caption{Normalised input spectrum for a Monte Carlo analysis 
running the experiment design algorithm, 
considering two input channels and 5 parameters. 
The solid line represents the mean, for each channel, 
of the 1000 runs while the grey shadow is the corresponding
standard variation. 
 \label{fig.robustness}}
\end{center}
\end{figure}

It is worth noticing here that the allowed range of discrepancy 
in the parameters is orders of magnitude higher than the expected 
uncertainty in the parameters. Indeed, previous studies~\cite{Nofrarias10,Ferraioli11,Congedo12, Karnesis14, Vitale14} have shown that the order of magnitude of expected error 
on the parameters considered in our analysis is 
$\sigma_{G_{\rm df}}  \simeq  \sigma_{G_{\rm lfs}} \simeq 10^{-5}$ for the control gains, 
$\sigma_{\omega^2_{1}}  \simeq  \sigma_{\omega^2_{2}} \simeq 10^{-10} \, \rm{s^{-2}}$ for the stiffnesses and 
$\sigma_{S_{21}} \simeq 10^{-8} $ for the sensing cross-coupling.

We run 1000 analysis of the algorithm described section ~\ref{sec.design}.
Each of the individual runs computes the dispersion function 
in a 100 frequency bins grid spanning from 0.1\,mHz to 1Hz. 
The input spectrum is built iteratively using the 
dispersion function as figure of merit. 
Each individual input spectrum 
of the Monte Carlo analysis was computed based 
on 50 iterations of the algorithm.
The result of the Monte Carlo run is shown in Fig.~\ref{fig.robustness}. 
The solid line represents, for each channel, 
the mean of all the input spectrum while the shadow area 
stands for the standard deviation from all the runs. 
As clearly shown, in the LISA Pathfinder mission framework 
and considering a scenario of high uncertainty in the parameters,  
the methodology described is still robust. 
Results show a clear convergence around a input power spectrum
for both channels and guarantee that in a mission 
realistic scenario ---with unknown parameter values---  
the methodology described here is a valid protocol
to define the calibration signals that will help us get the
maximum information from the experiments and, hence,
to optimise the mission timeline.

\section{Conclusions}

LISA Pathfinder and future space-borne gravitational wave detectors will require 
precise calibration of their dynamical systems in order to operate at their design sensitivities. Given the operational constraints for such missions, the design of 
injection signals used for calibration is a key aspect for efficient characterisation
of the instrument.
 
We have introduced a methodology to design experiments 
for these instruments based on the minimisation of 
the uncertainty ellipsoid in parameter space. This methodology
allows one to decompose the Fisher information matrix in its different contributions,
each related to a unique physical coupling ---or transfer function--- of the experiment. 
By studying these contributions we can evaluate the expected error for a 
given spectrum of the injected test signal. 
 
We have compared this with a numerical algorithm capable of 
generating an optimal input signal by iteratively improving a
proposed input spectrum. The algorithm uses the dispersion function
of the system to promote those frequencies which minimise
the error on the parameters under study. We have applied both techniques to one example of LISA Pathfinder injection experiments, obtaining agreement in the injection signals obtained with both approaches.

As an example, we have 
considered the contributions to the expected error for a given term of the 
Fisher matrix decomposition: the $F_{2121}$, which describes the coupling of the 
$x_1$ (the drag-free channel) and the $x_{12}$ (the differential channel) for the case when a signal is injected in the former. The analysis is however general and can be readily extended to other experiments within LISA Pathfinder.

The methodology proposed here is general and can be equally applied to 
other instruments requiring an accurate calibration in terms of parameters uncertainties,
such as ground-based gravitational wave detectors.

%The fact that the signal is derived 
%starting from a flat spectrum implies that the input signal obtained at the end
%of the analysis is basically derived in terms of the sensitivity of the experiment 
%to the given parameters that we use to describe it. 
%
%We showed as well that the formalism can be used to determine the amplitude 
%of the input signals, in the case that the frequencies of the injected signals are already chosen.
%
%It is important to notice that the current version of the technical note is focused on
%establishing a well based methodology.
%
% The results obtained in terms of the expected
%error on parameters should not be considered true predictions until the
%viabilty of the proposed signals is considered in more detail. Once a maximum for
%the allowed signal is set, it will directly rescale the errors on the parameters
%presented in this technical note.

%\input{include/appendix}
%^\input{include/tablefig}


\bibliographystyle{apsrev4-1}
\bibliography{library}

\end{document}